\documentclass[twocolumn,showpacs,amsmath,amssymb,pra,superscriptaddress,nofootinbib]{revtex4}

\usepackage{graphicx}

\usepackage{amsmath}

\newcommand{\bra}[1]{\mbox{$\left\langle #1 \right|$}}
\newcommand{\ket}[1]{\mbox{$\left| #1 \right\rangle$}}

\begin{document}

\title{Detector-decoy quantum key distribution without monitoring signal disturbance}

\author{Hua-Lei Yin}\email{hlyin@mail.ustc.edu.cn}
\author{Yao Fu}\email{yaofu@mail.ustc.edu.cn}
\author{Yingqiu Mao}
\author{Zeng-Bing Chen}\email{zbchen@ustc.edu.cn}
\affiliation{Hefei National Laboratory for Physical Sciences at Microscale and Department of Modern Physics,\\University of Science and Technology of China, Hefei, Anhui 230026, China}
\affiliation{The CAS Center for Excellence in QIQP and the Synergetic Innovation Center for QIQP, \\
University of Science and Technology of China, Hefei, Anhui 230026, China \\}

\begin{abstract}
The round-robin differential phase-shift quantum key distribution protocol provides a secure way to exchange private information without monitoring conventional disturbances and still maintains a high tolerance of noise, making it desirable for practical implementations of quantum key distribution. However, photon number resolving detectors are required to ensure that the detected signals are single photons in the original protocol. Here, we adopt the detector-decoy method and give the bounds to the fraction of detected events from single photons. Utilizing the advantages of the protocol, we provide a practical method of performing the protocol with desirable performances requiring only threshold single-photon detectors.
\end{abstract}

\maketitle

\section{Introduction}

Quantum key distribution (QKD) allows two legitimate users, typically called Alice and Bob, to share a common bit string with information-theoretic security even in the presence of eavesdroppers \cite{BB_84,ekert1991quantum}. Since the BB84 \cite{BB_84} protocol was proposed, tremendous progress has been made, for example, SARG04 QKD \cite{SARG04:Quantum}, decoy-state QKD \cite{Lo:2005:Decoy,Wang:2005:Beating}, measurement-device-independent QKD \cite{Lo:2012:MDI,Braunstein:2012:MDI}, and device-independent QKD \cite{Acin:2007:Device}, were proposed to enhance the security of quantum communication. The security proof of QKD, which leads to the explicit form of the extractable secure key rate the corresponding protocol provides, is closely related to the original version of Heisenberg's uncertainty principle \cite{RevModPhys:09:Scarani}. It means that any eavesdropper's intervention acquiring the effective private information of quantum states will lead to a disturbance which can be discovered and estimated from a randomly chosen portion of measurement results, namely, monitoring the signal disturbance. The more disturbance that the eavesdropper (Eve) should have caused, the less efficient the QKD protocol will be.

Recently, Sasaki, Yamamoto, and Koashi proposed a ground-breaking approach, a qudit-based protocol, i.e., the round-robin differential phase-shift QKD (RRDPS-QKD) protocol that does not require disturbance monitoring since the limit on leaked information, namely, the portion of the sifted key subjected to privacy amplification, can be acquired in advance and maintains a high tolerance of noise \cite{Sasaki:2014:RRDPS}.
Since the RRDPS-QKD was proposed, it has been studied both theoretically \cite{Guan:2015:PRRDPS,zhang:2015:round,mizutani:2015:robustness} and experimentally  \cite{Guan:2015:PRRDPS,takesue:2015:experimental,wang:2015experimental,li:2015:experimental}.
According to the original protocol, the phase error estimation can be done after Alice's preparation in advance without considering Eve's interventions, which makes it independent of the bit error rate and thus incredibly desirable for practical implementations of QKD.
Specifically, under ideal circumstances, the RRDPS-QKD protocol can tolerate a high bit error rate, up to almost $50\%$, which is significantly different from previous QKD protocols \cite{RevModPhys:09:Scarani}, such as the qubit-based BB84 protocol, whose bit error rate cannot go beyond $11\%$ based on one way classical post-processing \cite{Shor:2000:Simple}.
However, as pointed out by the authors, a realization of this protocol requires Bob to be equipped with photon number resolving detectors (PNR). This is a problem that all experimental implementations of QKD based on qubit encoding today face, the requirement of photon-counting techniques, as their unconditional security is based on single photon transmission, in which Alice sends single photons into insecure quantum channels, and Bob only receives single photons. Yet in practice, this assumption cannot be satisfied due to the fact that weak laser pulses are usually used as the source, which occasionally include more than one photon, and that the eavesdropper may intercept and send multiphotons to the receiver.
Therefore, for qubit-based quantum communication protocols \cite{BB_84,ekert1991quantum,SARG04:Quantum,Lo:2005:Decoy,Wang:2005:Beating,Lo:2012:MDI,Yin:2014:long,Ma:2012:Alternative,Fu:2015:MQC,Yu:2015:Statistical,yin:2015:practical}, the decoy state method \cite{Lo:2005:Decoy,Wang:2005:Beating} has solved the multiphoton problems at the source with great enhancements, while squash models \cite{GLLP:2004:Security,Beaudry:2008:Squashing,Fung:2011:squash,Gittsovich:2014:squash} are proposed to solve problems at the detector.

In the RRDPS-QKD protocol \cite{Sasaki:2014:RRDPS}, a practical vulnerability lies in that Bob's measurement device requires experimentally challenging detectors that are able to discriminate between single photons from two or more photons, i.e., photon number resolving detectors. Laboratories are presently equipped with conventional threshold photon detectors, and while actual PNR detectors are slowly entering commercial use, they are still highly temperature sensitive and can only resolve a limited number of photons received \cite{thomas2012practical}.
In fact, all experimental demonstrations of RRDPS-QKD have used threshold single-photon detectors \cite{Guan:2015:PRRDPS,takesue:2015:experimental,wang:2015experimental,li:2015:experimental}. Meanwhile, a recent work suggests that with the use of threshold detectors, security can still be achieved without monitoring the signal disturbance by employing a passive delay change at Bob's measurement site  \cite{Sasaki:2015:qcrypto}.
In this paper, we exploit a detector-decoy (DD) method that estimates the photon statistics provided by combining a threshold detector together with a variable attenuator (amplitude modulator) \cite{moroder2009detector} to give the bounds to the fraction of detected events by Bob from single photons. Through simulation and comparison with Ref.\cite{Sasaki:2015:qcrypto}, we show that with the photon statistics obtained, we have provided a more advantageous method for feasibly realizing the RRDPS-QKD with an enhancement in the key rate results.

\section{Method}
\subsection{RRDPS-QKD protocol}

\begin{figure}[t]
\centering
\resizebox{8cm}{!}{\includegraphics{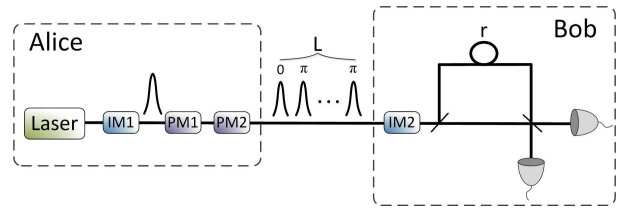}}
\caption{(Color online) Basic setup of a detector-decoy RRDPS-QKD. IM, intensity modulator; PM, phase modulator; r, the variable delay that generates random numbers from $0$ to $L-1$. Here, IM2 realizes Bob's detector-decoy method, PM1 adds a random phase on each pulse train, and PM2 encodes random phases $0$ or $\pi$ on each pulse.}
\label{f1}
\end{figure}

The basic procedures of the RRDPS-QKD protocol are as follows. First, Alice generates a series of pulse trains, each train with a overall random phase. Then, for each train, Alice prepares $L$ weak coherent pulses with the bit information encoded in their phases $s_{k}$,
\begin{eqnarray} \label{1}
\ket{\psi}=\bigotimes_{k=1}^L\ket{(-1)^{s_k}\alpha}=\bigotimes_{k=1}^L(-1)^{s_k \hat{n}_k}\ket{\alpha},
\end{eqnarray}
where $s_{k}\in \{0,1\}$, $\hat{n}_k$ is the photon number operator for the $k$th pulse, $\alpha$ is related to the average photon number per pulse with $|\alpha|^2=\mu/L$, where $\mu$ is the average photon number of each train. From there, Alice sends the quantum states to Bob through an insecure channel. At Bob's measurement site is an unbalanced Mach-Zehnder interferometer (MZI) with a variable delay at the long arm, which is controlled by a random number generator (RNG). He uses the RNG to generate a number $r\in \{1, ..., L-1\}$, and after some possible intervention from Eve, Bob detects the signal and acquires the indices $\{i,j\}$, where $i$ satisfies $j=i \pm r(\textrm{mod} L)$, and announces them via a public channel to Alice. In practice, the outputs of the MZI are adjusted so that superposed pulses of the same phase go to a detector $0$, while pulses of opposite phases go to a detector $1$. Thus, Alice records her sifted key as $s_{A}=s_{i} \oplus s_{j}$. A schematic diagram of the protocol is shown in Fig.\ref{f1}. Ideally, Bob's measurement outcomes $s_{B}$ should equal $s_{A}$, though in reality it may include some errors.
As Bob's random choice is after Eve's disturbance, the information leaked to Eve is very limited because of information causality \cite{Pawowski2009}. Intuitively, Eve seems to have some control over the generation of index $i$, though it was shown by a virtual measurement scheme proposed in Ref.\cite{Sasaki:2014:RRDPS} that her control is in fact rather limited. Therefore, it is possible to ignore the signal disturbance and analyze errors, namely the privacy amplification, based only on the outcomes of Alice and Bob. A crucial condition for the guarantee of this protocol is for Bob to only declare a detection event successful when one photon is exactly detected by his detector and no other detection occurs along the rest of the pulse. In the following, we show how the single photon detection condition is satisfied with our simple detector-decoy method under current technology.

\subsection{Detector-decoy method}

The detector-decoy method, termed by Moroder \emph{et al} \cite{moroder2009detector} to emphasize its connection and applicability in QKD, is outlined thus. Suppose an initial phase randomized signal state (written as a classical mixture of Fock states) of $\rho_{\textrm{in}}=\sum_{n=0}^{\infty}p_{n} \ket{n}\bra{n}$ \cite{Zhao:2008:UntrustedSource}, with $\sum_{n=0}^{\infty}p_{n}=1$ and $n$ being the photon number, passes through intensity modulator (IM) with a transmittance $\eta$, and is detected by a threshold detector. The detection operation can be characterized by two operators, receiving no photons that results in no clicks $F_{\textrm{vac}}(\eta)$, and receiving at least one photon giving exactly one click $F_{\textrm{click}}(\eta)$, where $F_{\textrm{vac}}(\eta)=\sum_{n=0}^{\infty}(1-\eta)^n \ket{n}\bra{n}$ and $F_{\textrm{click}}(\eta)=\openone-F_{\textrm{vac}}(\eta)$, $\openone$ is the unit operator. Therefore, the probability of receiving no clicks is $p_{\textrm{vac}}(\eta)=\textrm{Tr}[F_{\textrm{vac}}(\eta)\rho_{\textrm{in}}]=\sum_{n=0}^{\infty}(1-\eta)^n p_{n}$. Notice that if we were to variate the transmittance $\eta=\{\eta_{1},\eta_{2},\eta_{3},...,\eta_{M}\}$, in principle, we would be able to obtain a sufficient set of linear functions to solve the unknown parameters, the photon probabilities $p_n$, thereby attaining the received photon number statistics,

\begin{eqnarray}
\setcounter{equation}{2}
\label{2}
p_{\textrm{vac}}(\eta_{1}) &=& \sum\limits_{n=0}^{\infty}{(1-\eta_{1})^n p_{n}}\nonumber\\
&&\vdots\\
p_{\textrm{vac}}(\eta_{M}) &=& \sum\limits_{n=0}^{\infty}{(1-\eta_{M})^n p_{n}}\nonumber.
\end{eqnarray}

In an actual setting, for the detector's imperfections, such as finite detection efficiency $\eta_{d}$ and a dark count probability $\epsilon$, we modify the operator $F_{vac}(\eta)$ as $F_{\textrm{vac}}(\eta)=(1-\epsilon)\sum_{n=0}^{\infty}(1-\eta\eta_{d})^n \ket{n}\bra{n}$, and $p_{\textrm{vac}}(\eta)$ will take the form of $p_{\textrm{vac}}(\eta)=(1-\epsilon)\sum_{n=0}^{\infty}(1-\eta\eta_{d})^n p_n$. Following the ideal detector case, we can also vary the transmittance of the IM to obtain a set of linear functions to deduce the values of $p_n$ and thus gain the signal photon number statistics.

\subsection{Key rate}

With the detector-decoy method, we obtain the signal photon number statistics required for the RRDPS protocol key rate generation. In our simulation model, we vary the transmittance of Bob's detector three times, $\eta=\{\eta_1, \eta_2, \eta_3\}$, specifically $\eta_1=1, \eta_2=0.8, \eta_3=0.6$, so that the probabilities of receiving clicks can be written as
\begin{equation} \label{}
\begin{aligned}
T_1=1-(1-p_d)\sum_{i=0}^{10}(1-\eta_1\eta_d)^i p_{i},\\
T_2=1-(1-p_d)\sum_{i=0}^{10}(1-\eta_2\eta_d)^i p_{i}, \\
T_3=1-(1-p_d)\sum_{i=0}^{10}(1-\eta_3\eta_d)^i p_{i},
\end{aligned}
\end{equation}
with $p_d$ as the total dark count probability of each train and $\eta_d$ as the detector efficiency. Here, we assume that events in which the signals received by Bob that involve more than $10$ photons are highly improbable and thus ignored. Next, we calculate a related value, the rate of detection $Q_{k}$ can be directly measured experimentally. Following methods of the decoy state QKD \cite{Lo:2005:Decoy,Wang:2005:Beating}, we have
\begin{equation} \label{}
\begin{aligned}
Y_{i}^{k}&=1-(1-p_d)(1-\eta_t\eta_k\eta_d)^i,\\
Q_k&=\sum_{i=0}^{\infty}e^{-\mu}\frac{\mu^i}{i!}Y_{i}^{k}\\
&=1-(1-p_{d})e^{-\mu\eta_t\eta_k\eta_d},\\
\end{aligned}
\end{equation}
where $k=1,2,3$ and $\eta_t=10^{-\beta\textrm{d}/10}$ is the efficiency of transmission related to the transmission distance $\textrm{d}$ and $\beta$ is the channel loss rate of the fiber.
Clearly, with each attenuation of $\eta_k$, $Q_{k}$ should equal the corresponding probability $T_{k}$ of receiving a click in the detector, and therefore this will be used as constraints in subsequent calculations. In the RRDPS protocol, Bob only declares a detection event successful when single photons are registered by his PNR detector, and the multiphoton signals received are discarded. Here, we consider a more realistic scenario, in which despite signals of multiphotons created by the source or the eavesdropper, throughout transmission and detection, only one photon from each multiphoton signal pulse survives for registration at the detector, while all other photons are lost, mathematically put as a minimum of a single photon transmission probability function,
\begin{eqnarray} \label{G}
G=\sum_{n=0}^{10}n\eta_{d}(1-\eta_d)^{n-1}p_{n},
\end{eqnarray}
under the constraints that $Q_k=T_k$ for $k=1,2,3$, where the photon number probabilities $p_{n}$ satisfy $\sum_{n=0}^{10}p_{n}=1$ and $0\le p_{n} \le 1$.
Meanwhile, in general QKD protocols, the length of secure key $K_{1}$ is obtained after subtracting the bits used for error reconciliation and privacy amplification \cite{Sasaki:2014:RRDPS}, written as $K_{1}=N[1-fH_{\textrm{ER}}-H_{\textrm{PA}}]$, where $N$ is the length of the sifted key, $H_{\textrm{ER}}$ and $H_{\textrm{PA}}$ are the costs for error reconciliation and privacy amplification, respectively, and $f$ is the parameter related to the efficiency of the employed error correction code. In standard calculations, it holds that $H_{\textrm{ER}}=h(e_\textrm{b})$ and $H_{\textrm{PA}}=h(e_{\textrm{ph}})$, where $h(x)$ is the Shannon entropy $h(x)=-x\log_2x-(1-x)\log_2(1-x)$, $e_{\textrm{b}}$ and $e_{\textrm{ph}}$ are the bit error rate and phase error rate.
In our computations, we inspect specifically the key rate per pulse, with its formula written as \cite{Sasaki:2014:RRDPS}
\begin{equation} \label{Rate}
\begin{aligned}
K_{2}=\frac{1}{L}\Big(&G_{\textrm{min}}-Qfh(e_\textrm{b})-\big[e_{\textrm{src}}\\
&+(G_{\textrm{min}}-e_{\textrm{src}})h\big(\frac {v_{\textrm{th}}} {L-1}\big)\big]\Big),
\end{aligned}
\end{equation}
where $Q$ is the overall gain, 
$G_{\textrm{min}}$ is the lower bound of Eq.\eqref{G}, and $e_{\textrm{src}}$ is a constant associated with the probability of finding more than $v_{\textrm{th}}$ photons in each pulse, written as \cite{Sasaki:2014:RRDPS}
\begin{eqnarray} \label{5}
P(n>v_{\textrm{th}}) \le e_{\textrm{src}}=1-\sum_{i=0}^{v_{\textrm{th}}}e^{-\mu} \frac{\mu^i}{i!}.
\end{eqnarray}
As we can see, the second and third terms in Eq.\eqref{Rate} are the error correction and privacy amplification terms, respectively.

\section{Simulation Results}

\begin{figure}[htb]
\centering
\resizebox{8cm}{!}{\includegraphics{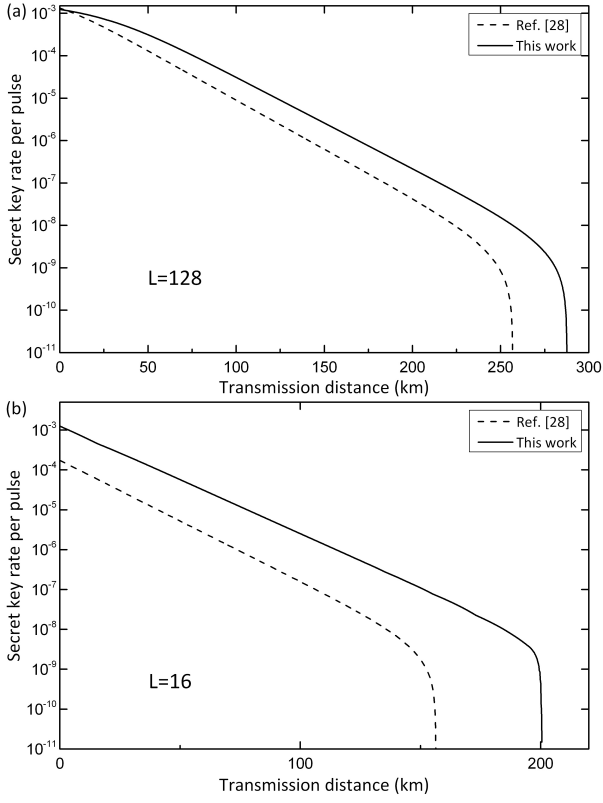}}
\caption{The optimized secret key rate per pulse for (a) $L=128$  and (b) $L=16$ in logarithmic scale as a function of the transmission distance. Our results (solid lines) show that after full optimization of the signal state $\mu$  and threshold photon number $v_{\textrm{th}}$ for each value of distance, the detector-decoy-based RRDPS-QKD gives higher optimal key rates and longer performance distances than the methods proposed in \cite{Sasaki:2015:qcrypto}, depicted as the dashed lines. In order to obtain nonzero key rates with our method at the transmission distance limit of $290$ km for $L=128$, $\mu$ and $v_{\textrm{th}}$ were optimized to $4.895$ and $20$ respectively, and $0.0535$ and $3$ respectively for $L=16$ at the transmission distance limit of $200$ km.}
\label{f2}
\end{figure}

In our simulation, the exact forms of $Q$ and $e_\textrm{b}$ in Eq.\eqref{Rate} can be given by \cite{Ma:2005:Practical}
\begin{equation} \label{}
\begin{aligned}
Q&=1-(1-p_d)e^{-\mu\eta_t\eta_d},\\
e_\textrm{b}&=[e_d(1-p_d)(1-e^{-\mu\eta_t\eta_d})+\frac{1}{2}p_d]/Q,
\end{aligned}
\end{equation}
where $e_d$ as the system error probability, and the RRDPS-QKD experimental parameters \cite{takesue:2015:experimental} used are listed in Table I.
\begin{table}[h]
\centering
\caption{Key parameters for simulation.}
\begin{tabular}{ccccc} \hline\hline
$p_d$ & $\eta_d$ & $e_d$ & $f$ & $\beta(\textrm{dB/km})$ \\ \hline
$1\times10^{-9}L$ & $19\%$ & $1.5\%$ & $1.16$ & $0.2$\\ \hline\hline
\end{tabular}
\end{table}

\begin{figure}[htb]
\centering
\resizebox{8cm}{!}{\includegraphics{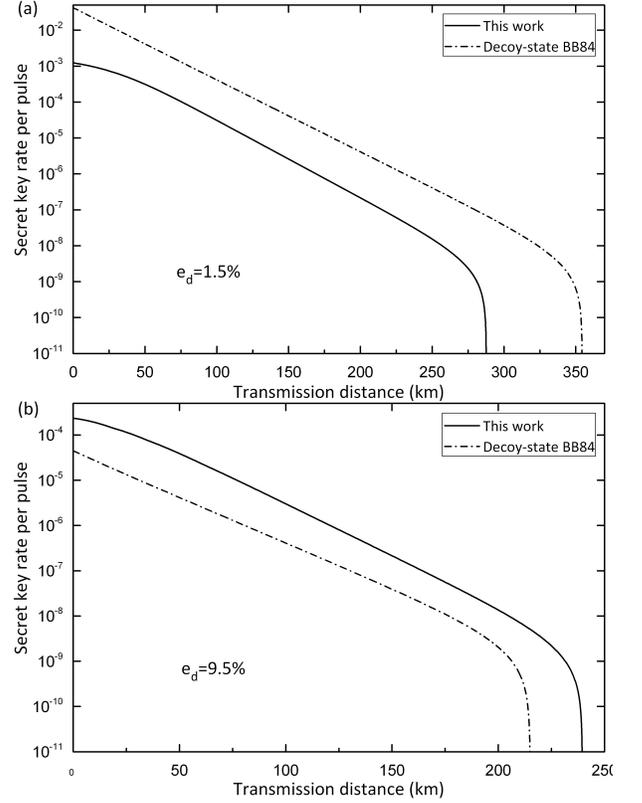}}
\caption{The optimized secret key rate per pulse for DD-RRDPS (solid lines) and BB84 with infinite decoy states \cite{Ma:2005:Practical} (dash-dotted lines) in logarithmic scale as a function of the transmission distance when $L=128$. (a) The key rates for $e_d=1.5\%$. (b) The key rates for $e_d=9.5\%$.}
\label{f3}
\end{figure}

Considering the worst case of the lower bound of $G$, where only one photon from each pulse of multiphoton signals survives transmission for registration at the detector, we optimize parameters $\mu$, the average photon number of each train, and $v_{\textrm{th}}$, the threshold photon number via a local search algorithm \cite{Xu:2014:Protocol} to obtain the optimal key rate per pulse $K_{2}$ as a function of transmission distance. The results are shown as the solid lines in Fig.\ref{f2}, for $L=128$ and $16$, from which we can see, our DD-RRDPS-QKD is an experimentally realizable protocol with desirable performance. We also give the full parameter optimized results of the recent implementation of RRDPS also using threshold detectors with a passive delay change \cite{Sasaki:2015:qcrypto} with the same simulation experimental parameters as our model (shown as the dashed lines in Fig.\ref{f2}) for comparison, where the key rate per pulse function $K_{3}$ is given as Eq.(2) of Ref.\cite{Sasaki:2015:qcrypto}
\begin{equation} \label{}
\begin{aligned}
K_{3}=\frac{1}{L}\Big(&Q-Qfh(e_\textrm{b})\\
&-\big[e_{\textrm{src}}+(Q-e_{\textrm{src}})h\big(\frac {2v_{\textrm{th}}} {L}\big)\big]\Big).
\end{aligned}
\end{equation}
As one can see, our results offer significant improvement in both key rate and transmission distance for a given pulse number $L$ compared with the methods proposed in \cite{Sasaki:2015:qcrypto}. Moreover, when $L$ becomes fewer, the advantages of our method become more prominent, as it offers more than one order of magnitude higher key rate per pulse at shorter distances to over two orders of magnitude higher key rate per pulse for longer distances, and greater performance distance of almost more than 50 km compared to \cite{Sasaki:2015:qcrypto}.

Furthermore, we give a comparison between our DD-RRDPS with conventional decoy-state BB84 \cite{Ma:2005:Practical}, shown in Fig.\ref{f3}. As one can see, when $e_d$ is small, the decoy-state BB84 outperforms the RRDPS, while as $e_d$ becomes larger, the advantages of RRDPS becomes more prominent. This can be explained by that while $e_d$ is small, $p_d$ plays a significant role in QBER, and since RRDPS encodes only one bit on $L$ pulses, the total detector dark count probability magnifies greatly and the average photon number per pulse decreases significantly, therefore limiting its transmission distance and secret key rate, whereas BB84 encodes one bit per pulse. However, when $e_d$ becomes very large, it will cost a very large portion of the key for privacy amplification in the BB84-QKD protocol, while for RRDPS-QKD, because of its high tolerance of errors, privacy amplification has nothing to do with the bit error rate. Therefore, the RRDPS greatly surpasses decoy-state BB84 given that the bit error rate is very large.

\section{Conclusion}
In conclusion, we have proposed a DD-RRDPS-QKD protocol that by using a threshold detector and a variable attenuator, a practical experimental implementation of RRDPS with desirable key rate and transmission distance has been achieved.
With our simulation results, we have given the bounds to the fraction of detected events by Bob from single photons and proved our work to be an experimentally realizable RRDPS protocol with better performance, even in the worst scenario of only one photon from the entire weak signal of multiphoton pulses is detected by the detector, and have shown that the results obtained in our protocol greatly surpasses recent works. Thus, an immediately feasible experimental solution of the RRDPS-QKD protocol under current technology is offered requiring only threshold single-photon detectors.

\acknowledgments
This work has been supported by the Chinese Academy of Sciences, the National Natural Science Foundation of China under Grant No. 61125502.



%

\end{document}